\documentclass[preprint,journal]{vgtc}       

\ifpdf
  \pdfoutput=1\relax                   
  \usepackage{graphicx}                
  \DeclareGraphicsExtensions{.pdf,.png,.jpg,.jpeg} 
\else
  \usepackage{graphicx}                
  \DeclareGraphicsExtensions{.eps}     
\fi%

\graphicspath{{figures/}{pictures/}{images/}{./}} 

\usepackage{microtype}                 
\PassOptionsToPackage{warn}{textcomp}  
\usepackage{textcomp}                  
\usepackage{mathptmx}                  
\usepackage{times}                     
\usepackage{cite}                      
\usepackage{tabu}                      
\usepackage{booktabs}                  
\usepackage[inline]{enumitem}
\usepackage{fixmath} 
\usepackage[autostyle]{csquotes} 
\usepackage{xcolor}
\usepackage[colorinlistoftodos]{todonotes}

\ieeedoi{10.1109/TVCG.2019.2934264}

\onlineid{1170}

\vgtccategory{Research}
\vgtcpapertype{theory/model}

\title{The Validity, Generalizability and Feasibility of Summative Evaluation Methods in Visual Analytics}

\author{Mosab Khayat, Morteza Karimzadeh, David S. Ebert, \textit{Fellow, IEEE}, Arif Ghafoor, \textit{Fellow, IEEE}}
\authorfooter{
\item Mosab Khayat, David S. Ebert, and Arif Ghafoor are with Purdue University. Email: \{mkhayat, ebertd, ghafoor\}@purdue.edu.
\item Morteza Karimzadeh is with the University of Colorado Boulder (formerly at Purdue University). Email: karimzadeh@colorado.edu.
}

\shortauthortitle{Khayat \MakeLowercase{\textit{et al.}}: An Analysis of Summative Evaluation of Visual Analytics Solutions}

\abstract{Many evaluation methods have been used to assess the usefulness of Visual Analytics (VA) solutions. These methods stem from a variety of origins with different assumptions and goals, which cause confusion about their proofing capabilities. Moreover, the lack of discussion about the evaluation processes may limit our potential to develop new evaluation methods specialized for VA. In this paper, we present an analysis of evaluation methods that have been used to summatively evaluate VA solutions. We provide a survey and taxonomy of the evaluation methods that have appeared in the VAST literature in the past two years. We then analyze these methods in terms of validity and generalizability of their findings, as well as the feasibility of using them. We propose a new metric called summative quality to compare evaluation methods according to their ability to prove usefulness, and make recommendations for selecting evaluation methods based on their summative quality in the VA domain.
} 


\CCScatlist{ 
 \CCScatTwelve{Human-centered computing}{Visu\-al\-iza\-tion}{Visualization design and evaluation methods}{}
}

\vgtcinsertpkg

\begin{document}


\firstsection{Introduction}

\maketitle


Visual analytics (VA) solutions emerged in the past decade and tackled many problems in a variety of domains. The power of combining the abilities of human and machine creates fertile ground for new solutions to grow. However, the rise of these hybrid solutions complicates the process of evaluation. Unlike automated algorithmic solutions, the behavior of visual analytics solutions depends on the user who operates them. This creates a new dimension of variability in the performance of the solutions that needs to be accounted for in evaluation. The existence of a human in the loop; however, allows researchers to borrow evaluation methods from other domains, such as sociology\cite{isenberg2008grounded}, to extract information with the help of the user. Such evaluation methods allow developers to assess their solutions even when a formal summative evaluation is not feasible. The challenge in these methods, however, lies in gathering and analyzing qualitative data to build valid evidence.

Many methods have been used to evaluate VA solutions, each originally developed to answer different questions with different evaluation intentions, including formative, summative and exploratory\cite{andrews2008evaluation}. Nevertheless, many of these methods have been extensively applied in summative evaluation despite the fact that some are only suitable for formative or exploratory assessment, not summative evaluation.

In this paper, we survey and analyze the evaluation methods commonly used with summative intentions in VA research. Specifically, we survey the papers presented at VAST 2017 and 2018, resulting in a seven-category taxonomy of evaluation methods. We identify the activities typically performed within each category, focusing on the activities that could introduce risks to the validity and the generalizability of the methods' findings, and we use both of these factors to define summative quality. We also define feasibility based on the identified activities and the limitations in applying evaluation methods in various scenarios. Finally, we use summative quality and feasibility to compare summative evaluation methods. Unlike existing problem-driven prescriptions \cite{munzner2009nested}, we analyze the risks to the validity, generalizability, and feasibility of each evaluation method by focusing on the activities employed in each method. We then provide a  prescription of evaluation methods based on (a) their ability to prove usefulness and (b) their feasibility.

The contributions of this paper can be summarized as follows:
\begin{itemize}
\item A survey, taxonomy and risk-based breakdown and analysis of evaluation methods used in summative evaluation of VA solutions.
\item A summative quality metric to assess the summative quality of evaluation methods based on the potential risks to the validity and the generalizability of the methods' findings.
\item An analysis and prescription of summative evaluation methods in terms of their summative quality and feasibility.
\end{itemize}

This paper is organized as follows: In Section 2, we provide important definitions, and review related work in Section 3. In Section 4, we present our survey of evaluation methods used for summative assessment. We analyze these methods in Section 5, followed by a set of recommendations for practitioners in Section 6. Finally, we conclude the paper and provide directions for future research.

\section{Usefulness and Summative Evaluation Definitions}\label{sec:Def}
The term \enquote{summative assessment} has roots in the field of education, which distinguishes it from formative assessment \cite{taras2005assessment}. The former assesses students objectively at the end of a study period using standardized exams, while the latter focuses on the learning process and the students' progress in meeting standards. In visualization and VA literature, summative evaluation has been traditionally referred to as the type of studies that measures the quality of a developed system using methods such as formal lab experiments \cite{munzner2009nested}. This is in contrast with formative assessment, which seeks to inform the design and development processes by applying techniques such as expert feedback \cite{nielsen1994heuristic}.

There have been some suggestions in the literature that evaluation intention should be unlinked from evaluation methods. Ellis and Dix \cite{EllisGeoffrey2006Aeao} argue that a formal lab experiment of a completely developed system can be conducted with formative intentions to suggest improvement. On the other hand, Munzner \cite{munzner2009nested} argues that formative methods such as expert feedback can be used with summative intentions to validate the outcome of different design stages. We agree with these arguments and believe that it is essential to give a formal definition of summative evaluation as an intention rather than an evaluation stage.

From the discussions in \cite{taras2005assessment}, we define summative evaluation as a systematic process which generates evidence about the degree of accomplishment of the given objectives (standards) for an assessed object (a solution) at a point in time. We use the term\enquote{solution} throughout this article to refer to different approaches for tackling a problem. VA research covers different types of solutions ranging from algorithms, to visualizations, to the integration of these in a holistic system (See \cite{sedlmair2016design} for design study contributions). \enquote{Standards} refer to benchmarks that are used to distinguish useful solutions from non-useful ones. These are commonly determined during the requirement elicitation stage, e.g. by conducting qualitative inquiries with domain experts. 

Summative evaluation is used to determine the usefulness (i.e. the value) of a solution. From a technology point of view, usefulness is based on two main factors: effectiveness and efficiency \cite{van2005value}. The former can be defined as the ability of a solution to accomplish the desired goals (i.e. doing the  right things). The latter concerns  the ability of a solution to optimize resources, such as time or cost, while performing its tasks (i.e. doing things right). Most existing summative evaluations assess one or both of these two factors.

Effectiveness and efficiency could be assessed differently according to the nature of the evaluated solution and the problem it tackles. Some solutions can be assessed in a straightforward manner because of the availability of explicit objectives they seek to achieve. An example of such solutions is a classification algorithm, which can be evaluated by objective metrics such as \textit{accuracy}. On the other hand, some solutions require extra effort to define valid objectives that can be used to assess their usefulness. Such effort can be seen in previous work targeted at finding valid objectives to determine the value of holistic visualization and visual analytics systems \cite{van2005value, stasko2014value, saraiya2005insight}.


Usefulness of human-in-the-loop solutions can also be assessed by utility and usability objectives. A Useful system has the needed functionalities (utility) designed in a manner that allows users to use them correctly (usability) \cite{nielsen1994usability}. The question of whether to prioritize utility or usability has been discussed in previous work \cite{grinstein2003comes}. We focus on the objectives used in utility and usability evaluation and view them with a broader lens as ways to assess effectiveness and efficiency.

We consider effectiveness and efficiency as generic objectives of summative evaluation. This permits us to put all the methods used to assess these two factors in the same plate and compare them in terms of the quality of their evidence and the feasibility of generating them.


\section{Related Work}\label{sec:relatedwork}

In this section, we review previous related work in three categories
\begin {enumerate*}[label=\itshape\alph*\upshape)]
\item surveys of evaluation practices, 
\item analysis of evaluation methodologies, and 
\item prescription of evaluation methods.
\end{enumerate*}

Multiple studies have surveyed existing evaluation practices. Lam \MakeLowercase{\textit{et al.}} \cite{lam2012empirical} suggest that it is reasonable to generate a taxonomy of evaluation studies by defining scenarios of evaluation practices that are common in the literature. Their extensive survey is unique and provides many insights for researchers. Specifically, seven scenarios of evaluation practices are discussed along with the goals of each, with examplar studies and methods used in each scenario. Isenberg \MakeLowercase{\textit{et al.}}\cite{isenberg2013systematic} continue this effort by extending the number of surveyed studies and introducing an eighth scenario of evaluation practices. These studies helped us build the backbone of our taxonomy as explained in Section \ref{sec:taxonomyMethodology}. The initial code to group evaluation methods in our survey was derived from Lam \MakeLowercase{\textit{et al.}} and Isenberg \MakeLowercase{\textit{et al.}}. We then gradually modified the coding of evaluation methods according to the studies we surveyed. In contrast to the grouping approach according to common evaluation practices taken by previous surveys, we focus on grouping evaluation methods based on  the similarities in each method's (sub)activities, with the ultimate goal of analyzing the potential risks associated with them, rather than simply describing the existing evaluation practices.

The next set of related work focuses on explaining and analyzing evaluation methodologies. Evaluation research in visualization and VA can be divided into two types from the perspective of human-involvement: human-dependent evaluation and human-independent evaluation. The methodology of the first type draws on behavioral and social science methodologies to study the effect of visual artifacts on the human operator. One of the most well-known taxonomies for classifying behavioral and social science methodologies that has been ported to the Human-Computer Interaction (HCI) community is proposed by McGrath \cite{mcgrath1995methodology}. This taxonomy was built based on the three main dimensions that any behavioral study seeks to maximize, which are
\begin {enumerate*}[label=\itshape\alph*\upshape)]
\item generalizability,
\item realism, and
\item precision.
\end{enumerate*}
Generalizability of a study determines the extent of applicability of the study findings to any observable cases in general. It is related to the concept of external validity of results. Realism is the representativeness of studied cases to situations that can be observed in the real world; i.e., it determines the level of ecological validity of the findings. Finally, the precision of a study measures the level of reliability and internal validity of the findings. McGrath argues that these dimensions cannot be maximized simultaneously, since increasing one adversely affects the others. He then reviews common methodologies in behavioral science and assigns them to a position in the space defined by the three dimensions. Our analysis of evaluation methods relies on many of the arguments made by McGrath. A key difference between our work and that of McGrath lies in the intention of targeted studies. Our work focuses on studies that have a summative intention of proving usefulness. Unlike the general view of McGrath's work, summative evaluation studies have unique characteristics that permit ranking according to the quality of proving usefulness, as we explain in Section \ref{sec:analysis}.

An early study that introduces McGrath's work to the information visualization evaluation context is done by Carpendale \cite{carpendale2008evaluating}, who provides a summary of different quantitative, qualitative and mixed methodologies along with a discussion about their limitations and challenges. A more recent work by Crisan and Elliott \cite{crisan2018evaluate} revisits quantitative, qualitative and mixed methodologies and provides guidance on when and how to correctly apply them. Instead of taking a general view of behavioral methodologies, we  use a unified lens to identify limitations in evaluation methods used to prove usefulness, which may follow different methodologies, but are indeed used with summative intentions. Similar to Crisan and Elliott, we use validity and generalizability as our analysis criteria and add the feasibility criterion to the analysis to determine the level of applicability of the methods.

The second type of evaluation in visualization and VA is human-independent. In this type of evaluation, researchers follow a quantitative methodology to assess visualization or VA systems without considering the human element. This includes computer science methods of evaluating automated algorithms \cite{cormen2009introduction} and  statistical methods for assessing machine learning models (e.g. \cite{kohavi1995study}). A unique quantitative methodology that has been used to evaluate visualization and VA solutions is the information theoretic framework proposed by Chen and Heike \cite{chen2010information}. This framework treats the pipeline of generating and consuming visual artifacts as a communication channel that communicates information from raw data, as the sender, to human perception as the receiver. Information theory framework has been used to define objective metrics such as the cost-benefit ratio \cite{chen2016may}, which has been recently used to build an ontological framework that supports the design and evaluation of VA systems \cite{chen2019AnOntologicalFramework}. We include human-independent methods in our analysis because they are summative by nature.

The last set of related work focuses on the prescription of evaluation methods by providing guidelines on what evaluation methods are suitable for different evaluation instances. Andrews \cite{andrews2008evaluation} proposes four evaluation stages during the development cycle of a system: \begin {enumerate*}[label=\itshape\alph*\upshape)]
\item before the design, 
\item before the implementation,
\item during implementation, and 
\item after implementation.
\end{enumerate*}
Andrews suggests that the purpose, as well as the method of evaluation, is defined by the stage. For example, evaluation studies conducted after the implementation are summative in purpose and usually use methods such as formal experiments or guideline scoring. A more sophisticated prescription of evaluation methods is proposed by Munzner \cite{munzner2009nested}, who defines four nested levels, each having a set of unique problems and tasks. During the design stage, developers face multiple problems on their way to the inner level, which requires validation of the design choices. After implementation, a sequence of validation must be performed at each level to validate the implementation on the way out of the nest. Munzner then prescribes different evaluation methods to be used in each validation step. Meyer \MakeLowercase{\textit{et al.}} \cite{meyer2012four} expand this model by focusing on each of the nested levels and proposing the concepts of blocks and guidelines. Blocks describe the outcomes of design studies at each level, and guidelines explain the relationship between blocks at the same level or across adjacent levels in the nest. Another extension to Munzner's work is  Mckenna \MakeLowercase{\textit{et al.}} \cite{mckenna2014design} who link the nested model to a general design activity framework. The framework breaks down the process of developing a visualization into four activities of understand, ideate, make and deploy.

One argument made by Munzner \cite{munzner2009nested} was the necessity of summative evaluation during each stage of design studies to evaluate the outcome of that individual stage. Sedlmair \MakeLowercase{\textit{et al.}} \cite{SedlmairM.2012DSMR} and Mckenna \MakeLowercase{\textit{et al.}} \cite{mckenna2014design} made similar arguments while describing the process of design studies. They make the case for considering non-quantitative methods, such as heuristic evaluation, for summative purposes. While the Munzner's nested model \cite{munzner2009nested} essentially prescribes evaluation methods based on the development stage, we focus our analysis and prescription based on the activities performed during evaluation, and judge the quality of evaluation findings (evidence of usefulness) based on the amount of risk introduced by the involved activities. Further, our approach adapts to different evaluation instances and prescribes relatively smaller number of potential evaluation methods, compared to \cite{munzner2009nested}.


Another form of prescription studies is the study of correctly adopting existing evaluation methods in the context of VA. Most evaluation methods that have been applied in visualization and VA have been borrowed from the field of human-computer interaction (HCI). Scholtz \cite{ScholtzJeanC.2018Ueov} explains the main factors that need to be added or modified in existing HCI methods to increase their utility in VA research. In addition, she prescribes potential evaluation metrics that have been successfully applied to assessing VA solutions. Still, the necessity of searching for suitable evaluation metrics for visual analytics persists \cite{ScholtzJeanC.2018Ueov,mahyar2015towards,kang2009evaluating,scholtz2006beyond}.

\section{Survey of Summative Evaluation Methods}\label{sec:Survey}

In this section, we present our survey and taxonomy of methods used by other researchers for the summative evaluation of VA solutions. Our goal in developing a taxonomy is to identify their limitations in terms of their validity, generalizability, and feasibility. Because of our objective of analyzing evaluation methods themselves, it is important to note that we abstract the evaluated solutions and the problems they solve. For example, we do not distinguish between a study that reports a holistic evaluation of a complete VA system and another study that evaluate a part of the system, as long as they both use the same evaluation method. This abstraction is discussed in Section \ref{sec:analysis}.

We focused our survey on papers that were published in VAST-17 and VAST-18. The initial number of papers we considered was 97 papers (52 papers from VAST17 and 45 papers from VAST18). We excluded papers that only included usage scenarios or did not report any evaluation at all. Usage scenarios are excluded since they only exemplify the utilization of solutions rather than systematically examining their usefulness. They differ from case studies and inspection methods, which have been used to systematically determining the usefulness of a solution as we explain next. The final number of papers we include in our taxonomy is 82. Some of these papers report more than one type of assessment. The total number of evaluation studies we found in these 82 papers is 182. The number of included papers are relatively small compared to existing surveys \cite{lam2012empirical, isenberg2013systematic}. However, our deductive approach to identify evaluation categories requires a smaller sample size compared to inductive approaches which develop concepts by grounding them to data. We built on previous taxonomies \cite{lam2012empirical,isenberg2013systematic} to layout ours, and then surveyed recent papers to guide the grouping, activity breakdown and risk analysis.


\subsection{Survey Methodology}\label{sec:taxonomyMethodology}

We followed a deductive approach to build our taxonomy, starting with an initial code based on the previous surveys \cite{lam2012empirical,isenberg2013systematic}, and then progressively changing the concepts in the code by considering new dimensions that help highlight factors that affect validity, generalizability, and feasibility of evaluation methods.

\paragraph{Phase 1: Building the initial concepts}
We based our taxonomy on two extensive surveys of evaluation practices in visualization and VA literature \cite{lam2012empirical,isenberg2013systematic}. The descriptive concepts developed in these works (i.e. the evaluation scenarios) are built for different objectives than our diagnostics. However, these works include the set of evaluation methods used in each scenario, which allowed us to determine our initial code. We consider each reported evaluation method as a concept in this phase and categorize the studies accordingly.

\begin{figure*}[th]
 \centering 
 \includegraphics[width=\linewidth]{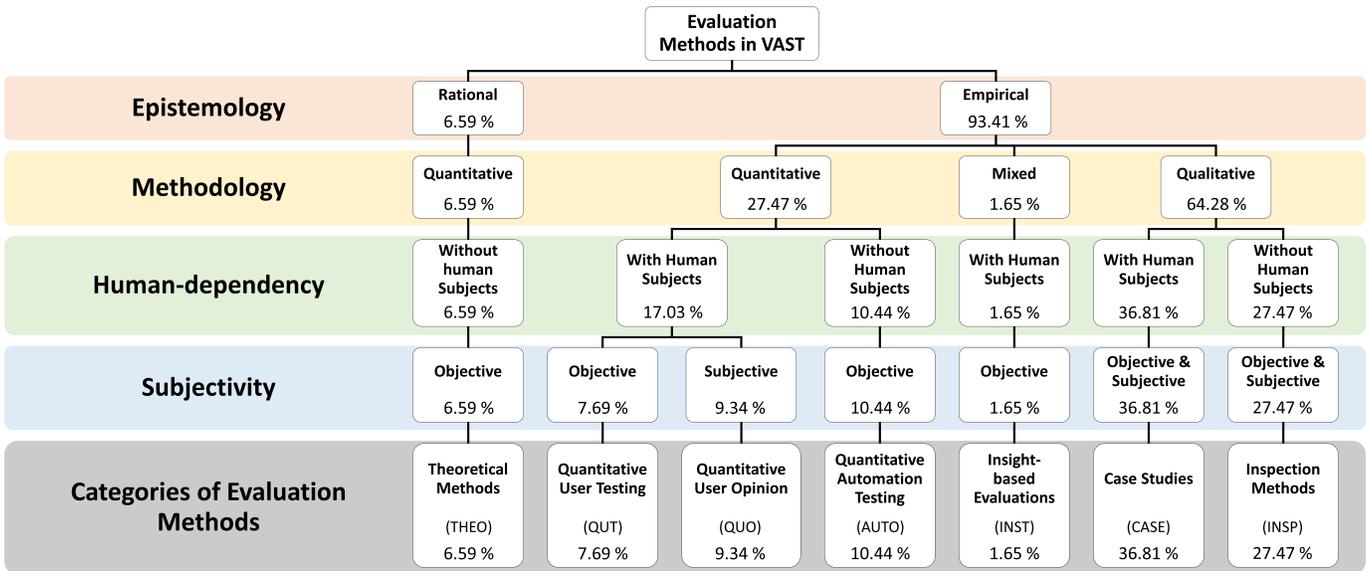}
 \caption{A taxonomy of summative evaluation methods based on surveying 82 papers published in VAST-17 and 18. The leaves represent categories of evaluation methods distinguished by the dimensions shown in the left. The percentages show the distribution of surveyed studies.}
 \label{fig:taxonomy}
\end{figure*}

\paragraph{Phase 2: Selecting grouping dimensions}
We looked for new dimensions that are key for diagnosing evaluation methods' validity, generalizability, and feasibility. By examining the process of evaluation in each method, we identified four dimensions that are useful in grouping the evaluation methods to simplify our analysis: epistemology, methodology, human-dependency, and subjectivity. These dimensions can be seen as titles for each level of our taxonomy depicted in Figure \ref{fig:taxonomy} and are explained in more detail in the following sections.

\paragraph{Phase 3: Redefining concepts}
We iteratively refined the taxonomy, which resulted in merging some concepts and splitting others. For example, one of our initial codes was \enquote{Quantitative-objective assessment} which included both \enquote{Quantitative User Testing} and \enquote{Quantitative Automation Testing} in our final code. The dimension responsible for splitting these two concepts is the \enquote{Human-dependency} dimension. On the other hand, we decided to merge \enquote{quantitative-subjective assessment test} and \enquote{quantitative-subjective comparison test} concepts into the single concept \enquote{Quantitative User Opinion}, because both concepts are similar in every grouping dimension that we considered.

\begin{table*}[tb]
  \caption{A summary of our survey of the evaluation studies reported in VAST-2017 and 2018. The table provides a brief description of our seven categories and the distribution of the surveyed studies within those categories. The total number of categorized studies is \textbf{182} reported in 82 papers.}
  \label{tab:evaluationCategory}
  \scriptsize%
    \centering%
    \begin{tabular}{m{0.04\linewidth}m{0.16\linewidth}m{0.45\linewidth}m{0.05\linewidth}m{0.08\linewidth}m{0.07\linewidth}}
  \toprule
Abb & Category &Description &\centering Frequency &\centering \% &\centering Examples\tabularnewline
  \midrule

\textbf{THEO} &Theoretical Methods & Rational, objective, quantitative methods which do not rely on human subjects to generate evidence of usefulness. These methods rely on deductive reasoning to logically derive evidence.&\centering 12 &\centering 6.59\% & \cite{andrienko2019analysis,chan2019v,chen2019information}\tabularnewline

\hline 

\textbf{QUT} & Quantitative User Testing& Empirical methods that are objective, quantitative and estimate the performance of human subjects for assessment or comparison reasons. &\centering 14 &\centering 7.69\% & \cite{zhao2019evaluating, angelini2019vulnus, liu2019interactive}\tabularnewline

\hline 

\textbf{QUO} &Quantitative User Opinion & Similar to the previous category; except, it assesses subjective aspects instead of measuring objective performance. A conventional method in this category is structured questionnaires which use measurable scales, e.g. Likert scale \cite{kaptein2010powerful}, to evaluate user satisfaction and opinion. &\centering 17 &\centering 9.34\% & \cite{sultanum2019doccurate,pienta2018vigor,law2019duet}\tabularnewline

\hline 

\textbf{AUTO} &Quantitative Automation Testing & Empirical methods used to quantitatively and objectively assess human-independent solutions such as machine learning models. This includes evaluation methods such as cross-validation and hold-out test set to predict the performance of supervised machine learning models \cite{kohavi1995study}. &\centering 19 &\centering 10.44\% & \cite{wilkinson2018visualizing,xie2019visual,wang2019dqnviz} \tabularnewline

\hline 

\textbf{INST} &Insight-based & A mixed method which relied on human subjects to qualitatively identify a set of insights that can be reached with the help of a solution. Insight-based methods map identified insights to measurable metrics, e.g. insights count, which are used for quantitative reasoning \cite{saraiya2005insight}. &\centering 3 &\centering 1.65\% & \cite{LeiteRogerA.2018EVAt,zhao2018supporting,chung2017cricto}\tabularnewline

\hline 

\textbf{CASE} &Case Studies & Qualitative methods which allow researchers to determine objective values and subjective opinions about the evaluated solution by interacting with human subjects who are typically domain experts. This category encompasses different variants of case studies including Pair analytics \cite{arias2011pair} and Multi-dimensional In-depth Long-term Case studies \enquote{MILC} \cite{shneiderman2006strategies}. &\centering 67 &\centering 36.81\% & \cite{stein2018bring,wang2019visual,OrbanDaniel2019DaTA}\tabularnewline

\hline 

\textbf{INSP} &Inspection Methods & Methods which assess objective or subjective potentials of a solution without testing or recruiting human subjects. Inspection methods help in checking the satisfaction of predefined requirements that characterize objective or subjective features needed in useful solutions \cite{nielsen1994heuristic,ScholtzJeanC.2018Ueov,scholtz2011developing,tory2005evaluating,allendoerfer2005adapting}. &\centering 50 &\centering 27.47\% & \cite{janicke2017interactive,wu2019forvizor,pezzotti2018deepeyes}\tabularnewline
  \bottomrule
  \end{tabular}%
\end{table*}

    \subsection{Taxonomy}
Figure \ref{fig:taxonomy} summarizes our taxonomy. The dimensions are independent and can be used separately to classify evaluation methods. Therefore, the order of dimensions in Figure \ref{fig:taxonomy} is not important. However, we chose to present a breakdown leading to our identified seven categories. In this section, we explain the dimensions that differentiate the seven categories of evaluation methods and the distribution of the surveyed papers in each level. The following section focuses on the analysis of the processes and activities in each method category.

		\subsubsection{Epistemology Dimension}
 Evaluation methods produce evidence that justifies our beliefs about the value of the evaluated solution. The process of justification in the evaluation methods can be categorized, according to epistemological views, into two classes: rational and empirical. Rational evaluation methods use deductive reasoning by relying on logically true premises. For example, the analysis of algorithms complexity as reported in \cite{lin2018rclens,andrienko2019analysis} is rational. This method is used to evaluate the efficiency of an algorithm by determining the time required to execute its instructions. Another rational method of evaluation is the information-theoretic framework \cite{chen2010information} that is used in \cite{chen2019information} to study the cost-benefit of visualization in a virtual environment. Both of these methods are built on top of a set of basic premises that are assumed to hold, such as the assumption of unit execution time per the algorithm's instruction and the axioms of probability, respectively.

Empirical evaluation methods, on the other hand, follow inductive reasoning by collecting and using practical evidence to justify the value of the solution. Most categories of evaluation methods are empirical. An example of an empirical method is the estimation of automated models' performance as reported in \cite{wilkinson2018visualizing,xie2019visual}. Such estimations are performed empirically by measuring the performance of a solution in a number of test cases.

Our survey shows that 12 out of 182 evaluation studies (6.59\%) were conducted using rational methods. Only one (0.58\%) of these 12 studies uses the information-theoretic framework. The other 11 studies (6.08\%) applied the traditional analysis of algorithms. Empirical evaluation methods are reported as the method of evaluation in the remaining 170 studies (93.41\%).

		\subsubsection{Methodology Dimension}
Evaluation methods are categorized, according to the methodology they follow, into three classes: quantitative, qualitative and mixed methods \cite{carpendale2008evaluating,crisan2018evaluate}. Quantitative methods rely on measurable variables to interpret the evaluated criteria. They collect data in the form of quantities and analyze it using statistical procedures to generalize their findings. The evidence generated by these methods has high precision but a narrow scope, i.e. rejection of a hypothesis by measuring particular metrics. Thus, these methods are preferable for problems that are well-abstracted to a set of measurable objectives. Controlled experiments are examples of quantitative methods, used extensively in comparative evaluations such as the studies reported in \cite{bernard2018comparing} and \cite{zhao2019evaluating}. These studies aim to justify the value of a solution by comparing it to counterpart solutions.

Qualitative methods, on the other hand, have fewer restrictions on the type of data that can be collected from a study. They evaluate the usefulness of solutions which tackle less abstract, concrete problems using data that is less precise but more descriptive, such as narratives, voice/screen recordings, and interaction logs.  Such data can be generated as a result of observation, or with the active participation of human subjects such as in interviews and self-reporting techniques. The case studies reported in \cite{stein2018bring} and \cite{wang2019visual} are examples of qualitative methods used with a summative intention.

Mixed methodology integrates both quantitative and qualitative methods to produce better comprehensive studies \cite{crisan2018evaluate}. The most common way of following this methodology is to perform multiple complementary studies that are independent but serve the same summative intention (called a convergence mixed method design \cite{creswell2017designing}). For example, the authors of \cite{ParkDeokgun2018CTVA} report a controlled experiment as well as a case study with domain experts used to evaluate ConceptVector, a VA system that guides users in building lexicons for custom concepts. The results of both studies can be compared to support each other in proving the value of ConceptVector. Another way of mixing quantitative and qualitative methods is to connect the two types of data prior to analysis such as in an insight-based evaluation method \cite{saraiya2005insight}. This method starts by collecting qualitative data in the form of written or self-reported insights, then transforms this data into quantity, e.g. insight count, for analysis, such as the evaluation reported in\cite{LeiteRogerA.2018EVAt}. Since our taxonomy categorizes evaluation methods at individual resolution, we only categorize methods which follow embedded and merging designs, e.g. the insight-based method, as mixed methods.

According to our survey,  62 studies (34.07\%) out of 182 were conducted using quantitative methods. 117 studies (64.29\%) were conducted using qualitative methods, and only 3 studies (1.65\%) were conducted using the mixed method. According to this, qualitative methods constitute the majority of evaluations in VAST-17 and 18. 21 out of 82 (25.61\%) apply the convergence mixed method design.

		\subsubsection{Human-dependency Dimension}
Visual analytics solutions combine both human and automated processes to tackle problems \cite{keim2008visual}. Researchers may evaluate different components independently. For example, researchers may evaluate the efficiency of an automated algorithm \cite{chen2018sequence,muthumanickam2019identification}, or inspect the requirements of a user interface \cite{strobelt2019s}. Another option is to assess human-related tasks such as estimating the performance of the users \cite{ming2019rulematrix} or gathering expert feedback about the value of a VA system holistically \cite{zhang2019manifold}.

The human-dependency dimension in our taxonomy affects all the factors we aim to analyze (i.e. validity, generalizability, and feasibility); therefore, we include it as a dimension in the taxonomy.

Our survey shows that 81 (44.51\%) out of 182 studies summatively evaluated a solution without utilizing any human subjects. Among these studies, 31 studies (17.03\%) used quantitative methods and 50 (27.47\%) used qualitative methods in the form of inspection. On the other hand, 101 studies out of 182 (55.49\%) used methods that rely on human subjects. This includes 31, 67, and 3 studies using quantitative, qualitative, and mixed methods respectively (17.03\%, 36.81\%, and 1.65\% respectively). We remind the reader that the word solution is an abstract concept, which can represent automated algorithms, user interfaces or a complete VA system.

		\subsubsection{Subjectivity Dimension}
The usefulness of a solution can be determined by assessing the objective level of accomplishments. However, the objectives are sometimes defined as abstract ideas that cannot be directly or independently assessed. For example, VA systems have a general objective of generating insights about available data \cite{keim2008visual}. Such an abstract objective may not always be assessable by defined measures. From another angle, a correlation between subjective assessment such as user satisfaction in information systems and the usefulness of these systems has been shown \cite{gelderman1998relation}. Therefore, researchers include subjective assessment methods as ways of determining a solution's usefulness. Subjective assessment can be performed quantitatively \cite{sultanum2019doccurate,pienta2018vigor} or qualitatively \cite{fu2018ancestral}, and can be done with the help of human subjects \cite{ZhaoXun2018SVAo} or through inspecting the design without relying on human subjects \cite{law2019maqui}. Qualitative methods have the flexibility to assess both objective and subjective aspects.

There is a clear difference between summative evaluation methods that use objective versus subjective scopes. Objective methods assess effectiveness and efficiency of a solution in tackling the targeted problem, whereas subjective methods assess factors that correlate with that solution's capabilities (indirect assessment of usefulness). This led us to include the subjectivity dimension in our taxonomy, to highlight the differences between objective and subjective categories in terms of validity, generalizability, and feasibility.

Our survey shows that 48 (26.37\%) studies out of 182  applied objective evaluation methods. 17 studies (9.34\%) applied subjective evaluation and 117 studies (64.29\%) applied qualitative methods that are not restricted to a narrow scope and can assess both objective and subjective aspects.

		\subsubsection{The Seven Categories of Summative Evaluation Methods}
Table \ref{tab:evaluationCategory} summarizes the surveyed evaluation studies in our seven categories of summative evaluation, fully listed in the supplementary material. The most reported evaluation category in VAST-17 and 18 is case studies, followed by the inspection category. These two types are used significantly more than other evaluation categories. The high feasibility of case studies and inspections could be the reason for their popularity, as we explain in Section \ref{sec:analysisResults}. On the other hand, the least utilized evaluation category is the insight-based methods. Many of the reported studies that capture subjects' insights do not perform the second stage of defining quantitative measures from captured insights, and thus, end up in the case studies category in our taxonomy.

\section{An Analysis of Summative Evaluation Methods} \label{sec:analysis}
We analyze the identified seven evaluation categories in terms of validity, generalizability and feasibility, in order to compare their capability of proving usefulness, which is the objective of summative evaluation. Some of these methods are originally designed to address different evaluation requirements, such as formative or exploratory questions. However, we include them here, since they have been used by others to prove usefulness. Our focus is to analyze the process of evaluation itself regardless of the type of solutions they evaluate.

\begin{table*}[tb]
  \caption{The source of validity, generalizability and feasibility risks encountered when conducting summative evaluation studies.}
  \label{tab:risksDescription}
  \scriptsize%
    \centering%
    \begin{tabular}{m{0.20\linewidth}m{0.14\linewidth}m{0.59\linewidth}}
  \toprule
  
 Activity &  Relevant categories &  Description of the Risk \tabularnewline
  \midrule
  
Defining the objectives and the objective metric(s) &
 \textbf{THEO}, \textbf{QUT}, \textbf{AUTO} &
Some tasks do not have a clear objective, e.g. exploratory tasks (feasibility risk).
\tabularnewline

\hline

Abstracting the evaluated solution by a formal language &
 \textbf{THEO}, \textbf{AUTO} &
Some solutions cannot be automated with our current knowledge, e.g. human-dependent solutions. (feasibility risk)
\tabularnewline

\hline 

Deductively inferring the performance of the evaluated solution using a formal system &
 \textbf{THEO} &
Building a new formal system requires extraordinary work and high abstraction skills. Reusing a formal system requires skills of mapping abstract problems and performing mathematical deduction. (feasibility risk)
\tabularnewline

\hline 

Sampling problem instance(s) &
 \textbf{QUT}, \textbf{QUO}, \textbf{AUTO}, \textbf{INST}, \textbf{CASE} &
Relying on unrepresentative problem instances. (validity, generalizability risk)
\tabularnewline

\hline 

Sampling human subject(s) &
 \textbf{QUT}, \textbf{QUO}, \textbf{INST}, \textbf{CASE} & 
Relying on unrepresentative target users. (validity, generalizability risk)
\tabularnewline

\hline

Sampling competing solution(s) &
 \textbf{QUT}, \textbf{QUO}, \textbf{AUTO}, \textbf{INST}, \textbf{CASE} & 
Bias in selecting competing solutions included in a comparative evaluation study.(validity, generalizability risk)
\tabularnewline

\hline 

Identifying the ground-truth &
 \textbf{QUT}, \textbf{AUTO} & 
Unavailable ground-truth for a representative number of problem instances.(feasibility risk)
\tabularnewline

\hline 

Organizing studied treatments &
 \textbf{QUT}, \textbf{QUO}, \textbf{INST} &
Fail to eliminate confounders. (validity, generalizability risk)
\tabularnewline

\hline 

Statistical testing &
 \textbf{QUT}, \textbf{QUO}, \textbf{AUTO}, \textbf{INST} &
A potential reduction to the risk as a result of testing the statistical significance of quantitative analysis findings. (validity, generalizability risk reduction)
\tabularnewline

\hline 

Qualitatively identifying insights &
 \textbf{INST} &
Subjects potential miss-reporting of reached insights / researcher potential miss-collecting of reached insights. (validity, generalizability risk)
\tabularnewline

\hline 

Defining quantity from insights&
 \textbf{INST} &
Defining a metric that do not reflect the value of solutions. (validity risk)
\tabularnewline

\hline 

Collecting and interpreting qualitative data &
  \textbf{CASE} &
Missing essential pieces of information / misinterpreting the value of a solution evaluated using collected information. (validity, generalizability risk)
\tabularnewline

\hline 

Identifying the requirements / heuristics sources &
 \textbf{INSP} &
Relying on a source which provides less than needed requirements/heuristics to distinguish a useful solution from another. (validity, generalizability risk)
\tabularnewline

\hline 

Requirements / heuristics elicitation &
 \textbf{INSP} &
Mis-eliciting requirements / heuristics from the identified source. (validity, generalizability risk)
\tabularnewline

\hline 

Judging the satisfaction of the requirements / heuristics &
 \textbf{INSP} &
Inspector subjectivity in checking the accomplishment of requirements / heuristics. (validity, generalizability risk)
\tabularnewline

\hline 

Indirect inference of usefulness &
 \textbf{QUO}, \textbf{CASE}, \textbf{INSP} &
Inferring the value of a solution from measures or findings that do not directly test the solution objectively. (validity, generalizability risk)
\tabularnewline

  \bottomrule
  \end{tabular}%
\end{table*}

	\subsection{Analysis Criteria} \label{sec:analysisCriteria}
Validity and generalizability are well-known properties of generated evidence in scientific studies and have been broken down into many types. The primary types influencing our analysis are internal validity and external validity as defined in experimental quantitative studies \cite{campbell1963experimental}, as well as credibility and transferability as defined in qualitative studies literature \cite{lincoln1985naturalistic}. We view validity as the property of correctness of study findings, while we see generalizability as the extent to which study findings can be applied to similar but unstudied (unevaluated) cases.

By examining the findings of each evaluation method, we found four types of summative evidence which assess effectiveness or efficiency:
\begin {enumerate}[label=\itshape\alph*\upshape)]
\item quantities that represent the objective performance (measured or estimated by a method from \textbf{THEO}, \textbf{QUT}, \textbf{AUTO}, or \textbf{INST}),
\item quantities that represent subjective satisfaction (estimated by a method from the \textbf{QUO} category),
\item qualitative information about objective or subjective value of a solution (gathered by \textbf{CASE} methods),
\item accomplishment of requirements/heuristics (inspected by a method belonging to \textbf{INSP} category).
\end{enumerate}
Each evaluation method includes a set of activities resulting in one of the aforementioned four types of evidence. In our analysis, we outline the activities for each method and highlight risk factors associated with each activity. We rely on the definition of risk found in the software engineering literature \cite{bannerman2008risk}, which defines exposure to risk as the probability-weighted impact of an event on a project (evaluation in our case). The identified risk factors may affect the validity and generalizability of the outcome of each method. For example, the generalizability of empirical evidence is affected by the sampling of cases for the study. Thus, in our analysis, we designate sampling as an activity for empirical evaluation methods and associate it with potential generalizability risk. On the contrary, some activities may reduce risks to validity or generalizability. For example, a typical activity to maintain the validity of quantitative empirical evidence is to apply inferential statistical tests \cite{marczyk2005essentials}. Such testing activity is an example of what we call a risk reducer.

Besides validity and generalizability, feasibility is the third criterion we consider in our analysis. We include this criterion to reason about researchers' decisions to evaluate solutions using methods with less summative quality. Table \ref{tab:risksDescription} describes the potential validity, generalizability and feasibility risks we identify for each of the summative evaluation category, along with the source of these risks.

	\subsection{Evaluation Process Breakdown} \label{sec:analysisResults}
\begin{figure*}[th]
 \centering 
 \includegraphics[width=\linewidth]{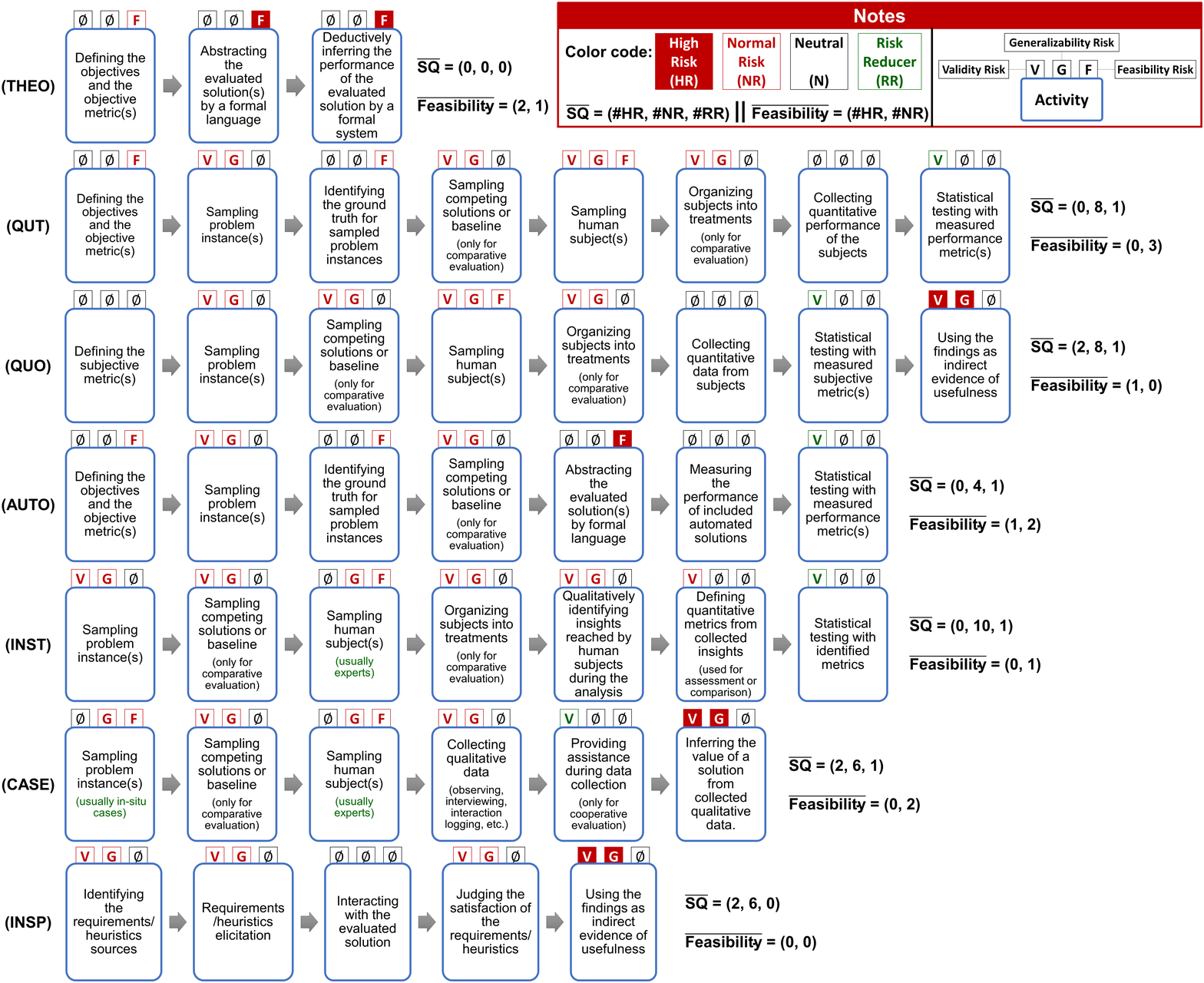}
 \caption{A summary of our analysis of evaluation methods. We capture the main activities taken by evaluation methods which could introduce risk to evidence validity, generalizability and feasibility. We assign 3 risk categories for these criteria per activity, classify each risk factor to high, normal or reducer class, then compare the methods using their summative quality (SQ) and feasibility.}
 \label{fig:Analysis}
\end{figure*}

 We break down the (sub)activities common to the methods in each category of our taxonomy. Then, we highlight the risks introduced or reduced as a result of performing these activities. The process of identifying the activities and highlighting their associated risks was performed based on our personal experience, validated and  by the survey we report in \ref{sec:Survey}. Figure \ref{fig:Analysis} presents a summary of our analysis, along with risks highlighted on each activity.

		\subsubsection{Theoretical Methods (THEO)}
This category includes complexity Analysis of Algorithms \& information-theoretic framework. These rational methods start by defining an objective metric, e.g time complexity, which is a useful measurement for assessment or comparison tasks. To measure the metric, researchers are required to abstract the behavior of the solution using a formal language, e.g. a programming language (Figure \ref{fig:Analysis}). This explicitly means full knowledge about the behavior of the solution. The last activity is to build a formal system, e.g. Turing machine \cite{copeland2004essential}, and use the premises in that system, e.g. unit execution time per instruction, to deductively measure the defined objective metric. Most rational studies captured in our survey apply the analysis of algorithm method to measure the time complexity of algorithms that are abstract by nature, and thus do not require the second activity. Moreover, the Turing machine is an applicable formal system that can be used to perform the deduction in this context. Another set of rational studies, which are more sophisticated, rely on information theory premises \cite{chen2016information}. Most remarkably, these works present an abstraction activity for solutions that are not abstract by nature \cite{chen2019information,chen2019AnOntologicalFramework}.

The three activities we report for rational methods do not introduce any risk to the validity and generalizability criteria. They are rigorous activities that always measure what they claim to measure. Rational methods also evaluate abstract problems and solutions with well-defined behavior, and thus are completely generalizable to any untested cases. For example, finding the worst case time complexity for an algorithm as $O(n)$ means no observable case of input size $n$ will ever take longer than linear execution time. 

The issue of rational methods appears in the feasibility criterion. The first feasibility risk is introduced by the first activity, which defines an objective metric. In many problems, the objective metric might not be feasibly defined. For example, the general goal of VA systems is to generate insights about data, a goal that may not be easily assessed by measurable factors. The second activity introduces much more sever risk to the feasibility. Abstracting the evaluated solution's behavior using a formal language requires sufficient knowledge about that solution's behavior, which may not be possible for some types of solutions.  For example, it is challenging to develop a formal language representation of human analytical processes, which practically limits the applicability of this type of evaluation on human-in-the-loop solutions. Since a human in these solutions controls their behavior, and that we cannot replace a human with a completely automated machine, it is not feasible to describe the human user’s behavior using a formal language. If the behavior of the solution cannot be abstracted, the third activity becomes infeasible since it cannot be performed in a formal manner without an abstract, well-defined solution. Moreover, building a formal system to deductively infer the performance of a solution is challenging and requires high abstracting skill.

		\subsubsection{Quantitative User Testing (QUT)}
In these empirical quantitative methods, researchers study human-in-the-loop solutions by either conducting a formal comparative experiment or measuring the performance of the solutions independently. The latter can be considered a special case of the former. These methods start by defining objective metrics, similar to rational methods. However, a typical activity in all empirical methods is to sample test cases. These cases are determined by sampling problem instances and human subjects. In comparative evaluation studies, the sampling of test cases includes the sampling of competing solutions. To objectively estimate the performance of the solution, researchers need to define ground truth for tested problem instances, which can be either sampled or synthesized \cite{whiting2008creating}. After sampling the test cases, researchers organize human subjects into groups (treatments) according to the study design. Two common designs include the within-subject (repeated measures) and between-subject (independent measures) designs. After organizing the study according to the selected design, researchers test human subjects with the sampled problems and collect quantitative measures of performance for each subject. These performance measurements can subsequently be analyzed per treatment using statistical tests (e.g. Analysis Of VAriance \enquote{ANOVA}). For assessment studies, statistics provide a confidence interval of the measured performance score for the solution. For comparative evaluation, the statistical tests ensure the significance of the difference between the performance of treatments. Some accuse such typical hypothesis testing methodology \cite{cumming2014new}. Nevertheless, Null-hypothesis significance testing (NHST) remains the most recognized methodology in quantitative scientific work.

The activities in the QUT category introduce risk to every criterion we analyze. A risk to the validity and generalizability criteria can be introduced as a result of sampling bias that excludes cases included in the study claim, sampling an insufficient number of cases to prove the claim, or failing to eliminate confounders when organizing treatments. The second risk can be reduced by applying a statistical test to show the potential of observing the findings for represented cases in general. The third risk is not a concern for assessment methods that do not generate evidence of usefulness as a result of comparing treatments.

The activities of QUT introduce risk to the feasibility criterion as well. Sampling representative cases can be infeasible because of the unavailability of representative human subjects or representative problem instances with known ground truth. Moreover, as in rational methods, it may not always be possible to identify a clear, objective metric that correctly distinguishes useful solutions from non-useful ones.
	
		\subsubsection{Quantitative User Opinion (QUO)}
The activities in this category are quite similar to the previous category. However, the focus here is on assessing subjective aspects instead of the objective performance, and there is no need to establish ground truth for the test cases.

The difference between subjective and objective methods, in terms of risk can be illustrated as follows. The risk to the validity is higher in subjective methods, since besides potential sampling and assignment biases, subjective methods do not assess usefulness directly. As we have mentioned earlier, the evaluation of usefulness by definition is a way to assess solutions objectively. Subjective methods approach achieve this by assessing factors that are assumed to correlate with usefulness, such as user satisfaction. However, such correlation may not always be valid. According to Nelson \cite{nielsen1994usability}, a system with limited utility could have high usability but would not be useful because of the missing functionalities. However, subjective methods are more feasible than objective methods. They do not require knowledge about ground truth nor quantifying objectives, and thus can be applied in more cases.
	
		\subsubsection{Quantitative Automation Testing (AUTO)}
These methods apply the same activities as \textbf{THEO} methods. The only difference between the two categories is the method of measuring the objective metrics for abstract solutions. In \textbf{THEO}, extensive work is devoted to building the formal system used in deduction, which is challenging because it requires high abstraction skills and sufficient knowledge about the problem domain. An alternative approach, taken by methods in the \textbf{AUTO} category, is to prove usefulness empirically by relying on sampled cases and statistics. For example, most methods used to evaluate machine learning models rely on estimating the performance with a set of testing problem instances \cite{kohavi1995study}.

The risk to the validity and generalizability of the evidence generated by a method from the \textbf{AUTO} category is slightly less than the risk associated with the \textbf{QUT} category. The reason is the reduction in sampling bias in \textbf{AUTO} methods as the result of excluding the human dimension. On the other hand, the exclusion of the human dimension explicitly means less feasibility of \textbf{AUTO} methods, since they are only capable of evaluating abstract solutions described by a formal language.
	
		\subsubsection{Insight-based Evaluations (INST)}
As an empirical category, sampling activities are typical in \textbf{INST}. A unique activity in this category is the qualitative data collection of insights. This is done by asking human subjects to self-report any insights they reach during the analysis by applying techniques such as diary \cite{saraiya2006insight} or think-aloud protocols \cite{smuc2009score}. Another unique activity in this category is the creation of measurable quantities out of collected qualitative data. The typical quantity to generate is insights count, which gives an indication of the usefulness of analytical support solutions.

Besides sampling bias, which can introduce risk to both validity and generalizability, \textbf{INST}'s unique activities may increase the risk to these criteria. For example, collecting insights as qualitative data introduces the possibility of misreporting some insights or misunderstanding reported ones. However, \textbf{INST} has a low feasibility risk since it does not require defining any objective metrics nor developing any tasks that ought to be evaluated quantitatively. \textbf{INST} also does not require prior knowledge about the ground truth of sampled problem instances.

		\subsubsection{Case Studies (CASE)}
Instead of measuring the accomplishment of solutions with some predefined metric (which may not be feasible or known for concrete domain problems), \textbf{CASE}  methods study realistic cases defined by actual real-world problem instances and intended users who are usually experts. To extract evidence of usefulness, evaluators pay extra attention to any data that can be captured during the examination. Collecting qualitative data is essential in case studies for creating a rich source of information, which helps in determining the usefulness of evaluated solutions. Many techniques can be implemented to generate qualitative data, including observation, semi-structured interviews, subject feedback, Think-aloud protocol, video/audio recordings, interaction logs, eye tracking and screen capturing \cite{carpendale2008evaluating}. During data collection, researchers may assist human subjects to overcome learnability issues. From the collected qualitative data, researchers can infer the value of evaluated solutions from the human subjects' perspective. This hypothesis of evaluated solutions' value can be used as evidence of usefulness, given that the human subjects are experts in the problem domain.

The risk to the validity and generalizability criteria for \textbf{CASE} methods can be explained as follows. Beside possible sampling bias, qualitative methods evaluate usefulness indirectly. The risk resulting from this indirectness can stem from two issues. The first is the potential misunderstanding of the human subjects when hypothesizing the value of the evaluated solution, which is typically known as the credibility of study findings. The second risk is the credibility of the subjects themselves, whose opinions are considered evidence of usefulness. This validity is affected primarily by how knowledgeable the subjects are about the problem domain, and secondarily by how much they know about using the evaluated solution. Another possible source of risk to the validity of case studies comes from the evaluators. The data collection and analysis in case studies can be profoundly affected by evaluators' subjectivity. Inexperienced evaluators may miss relevant information during data collection or wrongly infer the value of the solution from collected data. The risk introduced by the evaluators can be minimized by experience and by following guidelines that reduce subjectivity. There are tremendous existing literature on the correct application of qualitative studies \cite{strauss_corbin_1998,creswell_2018}. 

The advantage of case studies lies in their feasibility. They do not require specifying and measuring objective metrics or abstracting the solution. They also do not require knowledge about ground truth for the problem instances included in the test cases, because their objective assessment is derived from expert opinion, who are assumed to be capable of assessing the usefulness while testing the solution. The only feasibility risk to this category is the availability of expert human subjects, and the sampling of representative realistic problem instances.

		\subsubsection{Inspection Methods (INSP)}
The first activity of \textbf{INSP} is to identify the factors needed in useful systems through methods such as conducting a qualitative inquiry with stakeholders to identify requirements \cite{wu2019forvizor} or surveying the literature to identify known heuristics \cite{nielsen1994heuristic}. Once a set of requirements/heuristics is identified, researchers start inspecting the evaluated solution and judge whether it satisfies the identified requirements/heuristics.

 \textbf{INSP} includes the most feasible methods, not requiring human subjects nor testing with any problem instances. However, these methods prove usefulness marginally and with many validity and generalizability concerns. The risk to the validity and generalizability of the findings of \textbf{INSP} include (a) the credibility of the information source, (b) the exhaustiveness of the elicited requirements/heuristics, and (c) the subjectivity of the inspectors. Inspection methods have been shown to have significantly less potential for identifying usability issues compared to formal testing \cite{desurvire1994faster}. This finding inherently means high risks to both the validity and generalizability of  \textbf{INSP}'s evidence of usefulness.

	\subsection{A Ranking of the Summative Evaluation Categories}

After identifying risk factors to the validity and generalizability, we combine both criteria into a single metric which we call \textbf{summative quality} (SQ). The term is inspired by applied medical research for categorizing and ranking the quality of research evidence \cite{guyatt2008grade}. We define SQ as the probability of \textit{not} falling in any of the potential validity and generalizability risks introduced by a set of activities, i.e. the probability of an evidence to be valid and generalizable. Similarly, we consider feasibility as the probability of \textit{not} falling in any of the risk factors that threaten feasibility.

 SQ can be calculated by equation \ref{eq:summativeQuality}. We assume that the risks introduced by different activities are independent. Thus, to measure the total quality from subsequent activities, we take the product of the complement of the probability of risk in each activity. Taking the product is typical in similar total probability calculations (e.g. \cite{rohani2014calculating}). It is worth mentioning that the granularity of describing evaluation methods should not affect the total risk calculation. A single activity in a coarse-grained description of a method should accumulate all risk probabilities of that method when described in a fine-grained manner.

\begin{equation}\label{eq:summativeQuality}
SQ = \prod_{i=1}^{n} (1-P(risk_i))
\end{equation}

Equation \ref{eq:summativeQuality} measures the product of the probabilities of not falling in any of the $n$ validity and generalizability risks. This model of risk assessment requires estimating the probability of the captured risks, which is a challenging task. To overcome this issue and to be able to compare evaluation methods, we categorize the risk factors into three groups: high risk (HR), normal risk (NR) and risk reducers (RR) (Figure \ref{fig:Analysis}). High risk factors are introduced by any activities that infer usefulness indirectly (i.e., from evidence that do not measure objective metrics). Such activity would produce evidence of usefulness that have more uncertainty due to the high evaluators' potential subjectivity.

Using the categories of risk, we define $\overline{SQ}$ to compare evaluation methods In lieu of $SQ$. $\overline{SQ}$ can be defined as a triplet $\overline{SQ}=(\#HR,\#NR,\#RR)$, with each dimension representing the number of risk factors in each category. We calculate $\overline{SQ}$ for all categories then use the resultant triplets to observe any clear superiority of one category over another (e.g. (2,6,1) has less $\overline{SQ}$ given the two high risks compaerd to (0,8,1)). Based on this, we rank evaluation methods in terms of their $\overline{SQ}$ (Table \ref {tab:analysisScore}). The table also ranks evaluation methods based on  $\overline{feasibility}$, which can be defined as a tuple $(\#HR,\#NR)$ considering the feasibility risk factors. In case a clear superiority can not be decided (e.g. (0,10,1) Vs. (2,6,1)), we assign the same ranking to these methods (more examples of ranking calculations in the supplementary material). We stress that even though some categories rank low for $\overline{SQ}$, they may still be suitable for other purposes such as formative or exploratory.

\begin{table}[tb]
  \caption{The ranking of the seven categories of summative evaluation methods based on the potential risk to their validity, generalizability, and feasibility. We rank the categories according to their $\overline{SQ}$ and $\overline{feasibility}$.
}

  \label{tab:analysisScore}
  \scriptsize%
	\centering%
	\begin{tabular}{p{0.6cm}p{4.0cm}p{1.5cm}p{1cm}}
  \toprule
  \centering Abb & Category &\centering Summative Quality Rank &\centering Feasibility Rank \tabularnewline
  \midrule
  
\centering \textbf{THEO} & Theoretical Methods &\centering 1 &\centering 6
\tabularnewline

\rule{0pt}{2ex}\centering\textbf{QUT} & Quantitative User Testing &\centering 3 &\centering 4
\tabularnewline

\rule{0pt}{2ex}\centering \textbf{QUO} & Quantitative User Opinion &\centering 5 &\centering 2
\tabularnewline

\rule{0pt}{2ex}\centering \textbf{AUTO} & Quantitative Automation Testing &\centering 2&\centering 5
\tabularnewline

\rule{0pt}{2ex}\centering \textbf{INST} & Insight-based evaluation &\centering 4&\centering 2
\tabularnewline

\rule{0pt}{2ex}\centering \textbf{CASE} & Case studies &\centering 4&\centering 3
\tabularnewline

\rule{0pt}{2ex}\centering \textbf{INSP} & Inspection methods&\centering  4&\centering 1
\tabularnewline

  \bottomrule
  \end{tabular}%
\end{table}

\section{Recommendations}
Based on our taxonomy and analysis of summative evaluation methods, we provide the following recommendations:

\paragraph{1- Always select a feasible method with the highest summative quality.}

Prescribing an evaluation method for a given context can be done based on summative quality and feasibility. It is always encouraged to select the method with the highest summative quality. However, the feasibility of applying one of the methods in a given evaluation context may influence the selection. For example, the superiority of rational methods over empirical methods when testing usefulness; however, researchers may use an empirical method to evaluate a human-in-the-loop solution because of the infeasibility of abstracting human behavior using formal language as previously mentioned.

Our approach complements the nested model \cite{munzner2009nested}, which prescribes potential evaluation methods for each level. For instance, four different methods were prescribed to validate a solution in the encoding level. Complementing such prescriptions by following our approach can narrow down to a method from the Nested model prescribed methods.

\paragraph{2- Provide reasoning for evaluation method choice.}

We suggest providing solid reasoning when choosing an evaluation method for a summative evaluation. Our framework may help in this reasoning by considering the summative quality and feasibility as criteria. We note that it is always possible to use a weaker form of proving usefulness when it is feasible to generate stronger evidence with another method. For example, one can rely on subjective methods to assess the usefulness of a solution designed to tackle a problem that can be evaluated objectively. In such scenarios, evaluators should explain the limitation that prevents them from using the method that generates stronger evidence of usefulness. 

An example from the literature for a study that could have provided such an explanation is \cite{pezzotti2017deepeyes}. The authors used the inspection method to evaluate the usefulness of DeepEyes, a VA system developed to enhance designing deep neural networks. DeepEyes could have been evaluated using a formal controlled experiment i.e. by measuring training time and the classification accuracy of the end architecture (when using DeepEyes vs. traditional trial and error). Inspection has less summative quality compared to controlled experiments; thus, choosing the former over the later requires justification.

\paragraph{3- Encouraging insight-based evaluation.}

A surprising finding from our survey is the limited application of insight-based evaluation to published work in VA. According to our analysis, insight-based evaluation is one of the few methods that do not suffer from high risk factors. It is capable of assessing human analytical processes with realistic problems while generating quantitative outcomes that can be replicated and generalized. According to our survey, researchers favor case studies over insight-based methods in evaluation contexts that are suitable for both. We encourage performing insight  quantification and quantitative analysis instead of case studies to increase their precision and generalization potentials.

\paragraph{4- Apply multiple evaluation methods to minimize risk}

Our final recommendation encourages practitioners to apply multiple evaluation methods to prove the usefulness of their developed solutions. All of the evaluation methods include activities that could potentially invalidate the evidence they generate. An easy remedy is to compare the level of usefulness reached by different methods. This recommendation is strongly encouraged for subjective methods and inspection methods because of their relatively high validity and generalizability risks. Subjective methods are usually utilized to complement objective assessment, which is an excellent strategy for measuring usefulness from different angles.

\section{Conclusion and Future Work}
We presented our survey of evaluation practices used with summative intentions in VA. We identified seven categories of evaluation, broke down the activities in each, and analyzed each category in terms of feasibility as well as the validity and generalizability of their findings. We proposed summative quality as the primary metric for selecting evaluation methods for the summative intention of proving usefulness. Based on the summative quality metric and the complementary feasibility metric, we proposed a ranking of the categories of evaluation.

One of the limitations in our analysis is the possible subjectivity in identifying risk factors. We attempted to minimize it by continuously consulting the literature and conducting a survey. Assigning risk factors to only two categoris could also be considered a limitation. However, we favor robustness over precision when analyzing evaluation methods.

Even though we based risk analysis on extensive literature and our survey, our proposed ranking of evaluation methods might be considered subjective. Regardless, we argue that it characterizes the risks involved in selecting methods for summative evaluation, and most importantly, our risk analysis paves the way for future research and community ranking, similar to many repeated fruitful efforts in medical research \cite{guyatt2008grade}. 
Categorizing risks associated with activities and even quantifying such risks based on expert-assigned scores or probabilities is an established practice in system engineering and risk assessment\cite{haimes2015risk}, and our work lays the foundation for such analysis of evaluation methods in VA.

By identifying risk factors and providing a methodology, our work also enables community-driven prescription of evaluation methods. According to \cite{hubbard2010problems}, experts have high potential in judging risk factors and assigning probabilities. This approach can be used to assign probabilities to our identified risk factors using equation \ref{eq:summativeQuality} (see the supplementary materials for an example of such approach). To reduce subjectivity in judgment, one can deploy a community-driven voting system to increase the accuracy of estimating the risk probabilities and to build standards to prescribe evaluation methods.

Another direction to pursue in the future is to examine methods that have not been utilized in VA literature and their applicability in the field. An example of these methods is the formal specification and verification method \cite{bernot1991software}, which can evaluate the effectiveness of algorithms instead of measuring their efficiency. We are also interested in exploring mixed methods from other domains that resemble insight-based methods, because they can assess usefulness along with providing explanatory information to consider in relation to the captured performance.

\acknowledgments{
The authors wish to thank Christina Stober, Gourav Jhanwar, and  Bryan Jimenez for their comments and proofreading the paper.}

\bibliographystyle{abbrv-doi}

\bibliography{GEM_Draft8}
\end{document}